\documentclass[12pt]{iopart}
\usepackage{iopams}
\usepackage{setstack}
\usepackage{graphicx}
\usepackage{verbatim}
\usepackage{color}
\usepackage{subfigure}
\usepackage{hyperref}
\usepackage{xcolor}
\usepackage{dcolumn}
\usepackage{bm}
\usepackage{placeins}
\usepackage{enumitem}
\usepackage{siunitx}
\usepackage{hyperref}
\usepackage{cite}

\begin{document}
\title{Fabry-P\'erot~interferometry with gate-tunable 3D~topological~insulator nanowires}

\author{Javier~Osca$^{1,2}$\footnote{Present address:
Department of Theoretical Physics, Maynooth University, Ireland}, Kristof~Moors$^3$,
Bart~Sor\'ee$^{1,2,4}$ and Lloren\c{c}~Serra$^{5,6}$}
\address{$^1$ IMEC, Kapeldreef 75, B-3001 Leuven, Belgium}
\address{$^2$ KU Leuven, ESAT-MICAS, Kasteelpark Arenberg 10, 3001 Leuven, Belgium}
\address{$^3$ Peter Gr\"unberg Institut (PGI-9), Forschungszentrum J\"ulich, 52425 J\"ulich, Germany}
\address{$^4$ Universiteit Antwerpen, Departement Fysica, B-2020 Antwerpen, Belgium}
\address{$^5$ Institute of Interdisciplinary Physics and Complex Systems IFISC (CSIC-UIB), Palma, E-07122, Spain}
\address{$^6$ Department of Physics, University of the Balearic Islands, Palma, E-07122, Spain}
\ead{k.moors@fz-juelich.de, llorens.serra@uib.es}

\vspace{10pt}
\begin{indented}
\item[] \today
\end{indented}

\begin{abstract}
	Three-dimensional topological insulator (3D~TI) nanowires display remarkable magnetotransport properties that can be attributed to their spin-momentum-locked surface states such as quasiballistic transport and Aharonov-Bohm oscillations.
	Here, we focus on the transport properties of a 3D~TI nanowire with a gated section that forms an electronic Fabry-P\'erot (FP) interferometer that can be tuned to act as a surface-state filter or energy barrier.
	By tuning the carrier density and length of the gated section of the wire, the interference pattern can be controlled and the nanowire can become fully transparent for certain topological surface-state input modes while completely filtering out others.
	We also consider the interplay of FP interference with an external magnetic field, with which Klein tunneling can be induced, and transverse asymmetry of the gated section, e.g., due to a top-gated structure, which displays an interesting analogy with Rashba nanowires.
	Due to its rich conductance phenomenology, we propose a 3D~TI nanowire with gated section as an ideal setup for a detailed transport-based characterization of 3D~TI nanowire surface states near the Dirac point, which could be useful towards realizing 3D~TI nanowire-based topological superconductivity and Majorana bound states.
\end{abstract}


\submitto{\NT}

\maketitle


\section{Introduction} \label{Introduction}
A topological insulator (TI) material has the properties of a conventional insulator in the bulk but hosts topologically protected metallic states on the surface~\cite{TopoReview,Rew1,Rew2}.
The surface states have the peculiarity of behaving as massless Dirac fermions with a linear dispersion relation and a unique spin polarization that is tied to the momentum. The nontrivial topology of these TI materials is caused by strong spin-orbit coupling and a band inversion that leads to robust time-reversal symmetry protection of these surface states in the bulk gap.

Due to unintentional doping because of antisite defects~\cite{Hsieh919, Chuang2018}, for example, it is very challenging to fabricate 3D~TI samples with an intrinsic Fermi level that lies in the bulk gap and near the Dirac point. This makes it difficult to resolve the transport properties of 3D~TI surface states, with the majority of the charge carriers originating from the bulk.
To get into a regime where the 3D~TI surface-state (magneto)transport signatures are more pronounced, 3D~TI nanosamples with electrostatic gating are commonly considered~\cite{Xiu2011, Hong2014, Gooth2014, Jauregui2015, Cho2015, Jauregui2016, Ziegler2018, Muenning2021, Rosenbach2021}.

Usually, gated devices are considered in which the carrier density and Fermi level are shifted across the whole device, including the contact regions. Electrostatic gating that is restricted to a central section of a 3D~TI nanowire in between the contact regions offers additional advantages, however, especially when the length of the central section remains below the phase-coherence length and elastic mean free path of the surface-state charge carriers. In this case, the setup allows for resonant transmission of surface states across the gated section and electronic Fabry-P\'erot (FP) interferometry, a technique that is commonly considered in the context of fractional and integer quantum (spin) Hall effect edge channels~\cite{vanWees1989, Chamon1997, Camino2007, Zhang2009, Halperin2011, Dolcini2011, Citro2011, McClure2012, Romeo2012, Rizzo2013, Marciel2020}.

\begin{figure}[tb]
	\centering
	\includegraphics[width=\linewidth]{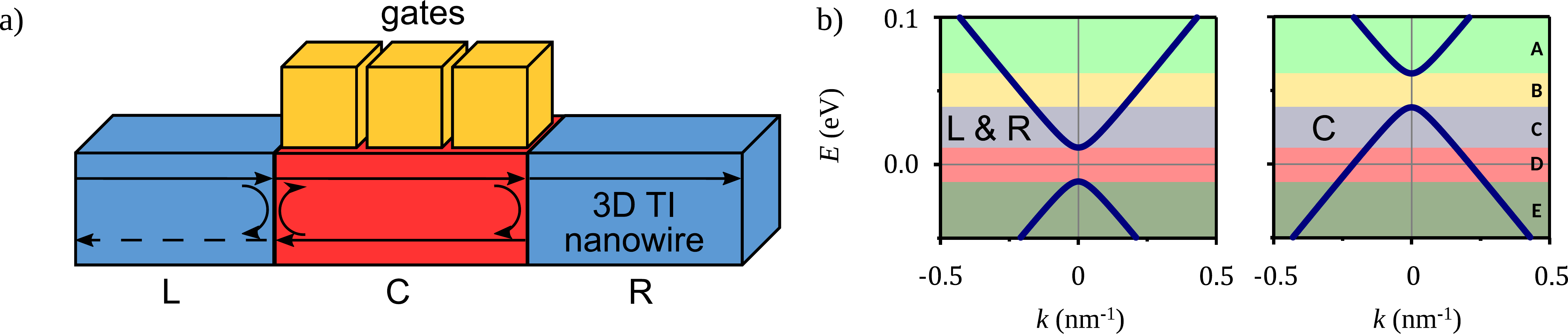}
	\caption{
		a) Device model schematic with surface-state transmission and reflection indicated (the dashed arrow for reflection indicating that perfect transmission can be realized due to Fabry-P\'erot interference). We consider a 3D~TI nanowire that is gated in its central section (dielectric not shown) such that the surface-state spectrum in this region can be shifted with respect to the nongated section on the left and right where the contacts are applied for the transport measurements. Multiple gates may be applied to control the length of the central section.
		b) A pair of surface-state (sub)bands ($j \neq 0$, $\nu = \pm 1$) of a cylindrical 3D~TI nanowire (model in section~\ref{Methods_2D}) with a gated central section. The spectrum in the leads (central section) is shown on the left (right), with a relative shift of $50\,\textnormal{meV}$. The qualitatively different transmission regimes, as discussed in section~\ref{Results_FP}, are presented in different colors and labeled from A to E.
	}
	\label{F1}
\end{figure}

In this paper, we focus on the calculation of the electronic transmission of surface states in a 3D~TI nanowire with a gated central section (see figure~\ref{F1}a). 
A capacitive gate is considered to locally shift the Fermi level of a central piece of the wire.
The surface-state conductance depends on the carrier density in the central section and on its length, which can be controlled by the gate voltage and by considering multiple (individually controllable) gates in sequence, respectively~\cite{Heedt2017}. 
By applying an external magnetic field along the wire, a perfectly transmitting mode is realized near the Dirac point~\cite{Bardarson2018}, which features Klein tunneling and is perfectly transmitted across the gated section, independent of the energy shift or length of the central section. We also consider the impact of transverse asymmetric gating, for realistic devices with a top gate, for example, on the transmission properties.

The paper is structured as follows. In the Methods section (section~\ref{Methods}), we present two different approaches for modeling the 3D~TI nanowire with gated section and its transmission properties. An overview of the transport properties of the setup under consideration is presented in section~\ref{Results_transmission} of the Results section (section~\ref{Results}), with subsections on FP interferometry, the impact of an external magnetic field, and transverse asymmetry of the gate effect, respectively. In section~\ref{Results_characterization} of the Results section, we expand on the experimental characterization of 3D~TI nanowire surface states by exploiting the transmission properties of the device under consideration. We discuss the results and draw conclusions in section~\ref{Discussion_Conclusion}.

\section{Methods} \label{Methods}
We consider two different models to describe the spin-momentum-locked surface states of 3D~TI nanowires. The first model considers the spinor solutions of the Rashba-Dirac Hamiltonian for the surface states of a cylindrical nanowire, which allows for analytical solutions for the transmission across a gated section.
The second approach is a numerical treatment based on a 3D continuum model for the band structure of 3D~TI materials that can be applied to nanowires with arbitrary cross sections with parameters that can be tailored to specific 3D~TI materials (e.g., describing their anisotropy and asymmetry between valence and conduction band). The two different models are summarized in the subsections below. Without loss of generality, we consider the left (right) lead to be the input (output) lead.

\subsection{2D surface-state model} \label{Methods_2D}
We consider the following Dirac delta-normalized plane-wave spinor eigenstates $\Psi_\nu(x, \phi)$ for the surface states of an infinitely extended cylindrical 3D~TI nanowire,
\numparts
\begin{eqnarray} \label{eq:2D_eigenstates_A}
	\Psi_\nu(x, \phi) &= \frac{1}{\sqrt{(2\pi)^{2}R}} \, e^{\mathrm{i} k x + \mathrm{i} l \phi} \, \Phi_\nu(l, k)\, , \\ \label{eq:2D_eigenstates_B}
	\Phi_\nu(l, k) &\equiv \left( \matrix{
		g_\nu(l, k) \cr
		e^{\mathrm{i} \phi} h_\nu(l, k) \cr
	} \right)\, ,
\end{eqnarray}
\endnumparts
with $\nu = +1$ ($\nu = -1$) for positive-energy (negative-energy) eigenstates that lie above (below) the Dirac point, $R$ the radius of the cross-sectional disc of the nanowire, $k$ the wave number along the direction of the wire, $l\in\mathbb{Z}$ the quantized orbital angular momentum along the circumference of the wire, $x$ the position coordinate along the wire direction, and $\phi$ the angular coordinate. The spinor components $g_\nu(l, k)$ and $h_\nu(l, k)$ are presented explicitly in \ref{A1}. Below, we refer to the positive-energy ($\nu = +1$) and negative-energy ($\nu = -1$) solutions as electron ($n$) and hole ($p$) modes, respectively.
The corresponding effective Hamiltonian for these spinor states is given by
\begin{equation} \label{eq:2D_Hamiltonian}
	H(j, k) = \hbar v_\textsc{f} \left( \matrix{
			-j/R & -\mathrm{i} k \cr
			\mathrm{i} k & j/R \cr
		} \right)\, ,
\end{equation}
with $v_\textsc{f}$ the Fermi velocity of the surface-state Dirac cone and $j \equiv l + 1/2 + \Phi/\Phi_0$ the generalized angular momentum containing the contribution of the Berry phase ($1/2$) and of the external magnetic field aligned with the nanowire. $\Phi/\Phi_0$ is the flux piercing the nanowire cross section in units of the flux quantum $\Phi_0 \equiv h/e(>0)$. The energies of the spinor solutions are given by
\begin{equation} \label{eq:spectrum}
	E_\nu(j, k) = E_\textsc{dp} + \nu \, \hbar \, v_\textsc{f} \sqrt{k^2 + j^2/R^2} \qquad (\nu = \pm 1)\, ,
\end{equation}
where $E_\textsc{dp}$ is the Dirac point energy (we consider $E_\textsc{dp} = 0$ below).

We calculate the transmission and reflection coefficients for a cylindrical 3D~TI nanowire with a central section where the surface-state spectrum is shifted by a constant of energy $\Delta E$, which is obtained by adding the spatial profile $\Delta E(x) = \Delta E \, [ \vartheta(x + L/2) - \vartheta(x - L/2)]$ to the Hamiltonian. The modes with different angular momenta decouple completely in the scattering problem and the transmission probability $T = |t|^2$ as well as 
the two-terminal
conductance $G = T e^2/h$ can be obtained analytically for the different input modes. More details can be found in \ref{A1}.

\subsection{3D~TI continuum model} \label{Methods_3D}
For 3D~TI nanowires with arbitrary cross sections and material-specific parameters, we consider the 3D continuum model Hamiltonian introduced in \cite{Zhang2,Zhang},
\begin{equation} \label{E1}
	\eqalign{H({\bf k}) &= \epsilon({\bf k}) + M({\bf k}) \tau_z \\
		&\hphantom{=}+ A_\perp \tau_x ( k_x \sigma_x + k_y \sigma_y) + A_\parallel \tau_x k_z \sigma_z\, ,}
\end{equation}
where
\begin{eqnarray}
	\epsilon({\bf k}) &=& C_0({\bf r}) + C_\perp (k_x^2 + k_y^2) - C_\parallel k_z^2\, ,
\label{E2} \\
	M({\bf k}) &=& M_0 + M_\perp (k_x^2 + k_y^2) - M_\parallel k_z^2\, .
\label{E3}
\end{eqnarray}
Here, $\tau_{x,y,z}$ and $\sigma_{x,y,z}$ are Pauli matrices for atomic-orbital and spin degrees of freedom, respectively. The results being presented here are obtained with the parameter set of \cite{Kristof}, representing an isotropic 3D~TI material with a bulk gap of $0.6\,\textnormal{eV}$, Dirac point of the surface states in the middle of the gap, and a Fermi velocity equal to $3\,\textnormal{eV}\cdot\si{\angstrom}/\hbar \approx 4.6 \times 10^5\,\textnormal{m/s}$: $M_0 = 0.3\,\textnormal{eV}$, $M_\perp = M_\parallel = 15\,\textnormal{eV}\cdot\si{\angstrom}^2$, $A_\perp = A_\parallel = 3\,\textnormal{eV}\cdot\si{\angstrom}$, $C_\perp = C_\parallel = 0$.

The three different regions of the 3D~TI nanowire with gated section are obtained by considering a piece-wise constant profile of $C_0$ along the wire direction ($x$), with $C_0 = 0$ in the leads (such that the Dirac point energy corresponds to $E = 0$, as in the 2D surface-state model) and $C_0 = \Delta E$ in the central section.

With this model and arbitrary wire cross sections, we need to resort to a numerical approach for obtaining the surface-state solutions in the different regions and the transmission coefficients. We use the complex band structure approach~\cite{Serra13, Osca2}, a method that is particularly well suited for our purposes since the computational demand does not increase with the length of the central region and only depends on the number of complex wave numbers included in each region and on the $yz$ grid at the two interfaces.

We expand the wave function in region $c = \mathrm{L},\mathrm{C},\mathrm{R}$ in its corresponding set of complex modes $\{k^{(c)}, \phi_k^{(c)}\}$, 
\begin{equation} \label{E26}
	\Psi^{(c)}(xyz,\sigma\tau) = \sum_k {d^{(c)}_k \, \exp{\left( \mathrm{i} k^{(c)} x \right)} \, \phi^{(c)}_k(yz,\sigma\tau) },
\end{equation}
where in $\mathrm{L}$ and $\mathrm{R}$ the amplitudes $d^{(c)}_k$ are classified as input or output depending on the mode flux. In $\mathrm{C}$, all modes are classified as output. If present, propagating modes are characterized by purely real wave numbers, while modes with $\mathrm{Im}{(k^{(c)})} \neq 0$ describe evanescent wave behavior emanating from the interfaces.
In (\ref{E26}), the components of the two pseudospins (one actual spin and one originating from the different atomic orbitals) are labeled by ${\sigma,\tau} = 1, 2$.

For a given energy $E$, the method proceeds in two steps. First, the three sets $\{k^{(c)}, \phi_k^{(c)}\}$ for the three different sections of the wire are obtained by solving independent eigenvalue problems in each section (see \ref{A2}). Second, a linear system determines the output amplitudes $d^{(c)}_k$ for a given input mode $k'$, with the corresponding transmission probability $T_{kk'} = |d^{(c)}_{k}|^2$. The linear conductance is finally given by the sum of transmissions from all (propagating) input modes of lead $\mathrm{L}$ to all output modes of lead $\mathrm{R}$: $G = (e^2/h) \sum_{kk'} T_{kk'}$.
This method has been reviewed recently in \cite{Osca2}, and it was used previously by some of us in \cite{Osca1, Osca3}.

\section{Results} \label{Results}

\subsection{3D~TI surface-state transmission} \label{Results_transmission}

\subsubsection{Fabry-P\'erot interference} \label{Results_FP}
The phase-coherent propagation of surface-state modes in the central section gives rise to an interference pattern in the overall transmission as a function of the barrier length or the wave number of the propagating modes in the central section, also known as electronic FP interferometry. Based on the results of the analytical model (details in \ref{A1}), we expect conductance oscillations for the different surface-state modes. These oscillations alternate between perfect transmission (i.e., when the reflection coefficient vanishes completely due to destructive interference) and a certain minimum, with an oscillation period given by $\Delta k_\textsc{c} L = 2 \pi$, with $L$ the length of the gated central section and $\Delta k_\textsc{c} = 2 |k_\textsc{c}|$ the wave number difference between forward- and backward-propagating modes in the central section with wave number $\pm k_\textsc{c}$. As the surface-state spectrum of 3D~TI nanowires is gapped [see (\ref{eq:spectrum}) and figure~\ref{F1}], the central section can also be tuned to a regime with evanescent modes, for which the transmission will be exponentially suppressed when increasing the length of the central section.
This surface-state gap is due to the transverse confinement of the 3D TI nanowire and, therefore, is 
inversely proportional to its transverse size (perimeter of the cross section).

Depending on the relative position of the surface-state spectrum in the leads and the central section (shifted by the gate effect), we can distinguish qualitatively different transmission regimes for each input mode (see figure~\ref{F1}b). Here, we consider an input mode 
in the left lead
with energy above the Dirac point energy (an electron mode) and a spectrum that is shifted upward in the central section with respect to the leads (by applying a negative gate voltage). 
As discussed below in section~\ref{Results_characterization},
there is a direct relation between the local shift of the bands and the gate potential, but here we simply assume a given
constant shift $\Delta E=50\; {\rm meV}$ in region C.
The bias potentials applied to L and R leads are very small,
infinitessimal in linear transport, and not affecting the
mode energies. They only cause an imbalance in  
occupations at the Fermi energy and a corresponding net 
L-to-R transmission.
In figure~\ref{F1}b we distinguish the following five transmission regimes that apply to different energy windows of the input mode and are labeled from A to E: (A) propagating electron modes in leads and central section (an $nnn$-junction), (B) propagating electron modes in leads and evanescent modes in central section (an $n\mathrm{b}n$-junction, with $\mathrm{b}$ referring to barrier, as the central section acts as a barrier in this case), (C) propagating electron modes in leads and propagating hole modes in central section (an $npn$-junction), (D) no input mode available, and (E) propagating hole modes in leads and central section (a $ppp$-junction).

\begin{figure}[tb]
	\centering
	\includegraphics[width=0.55\linewidth]{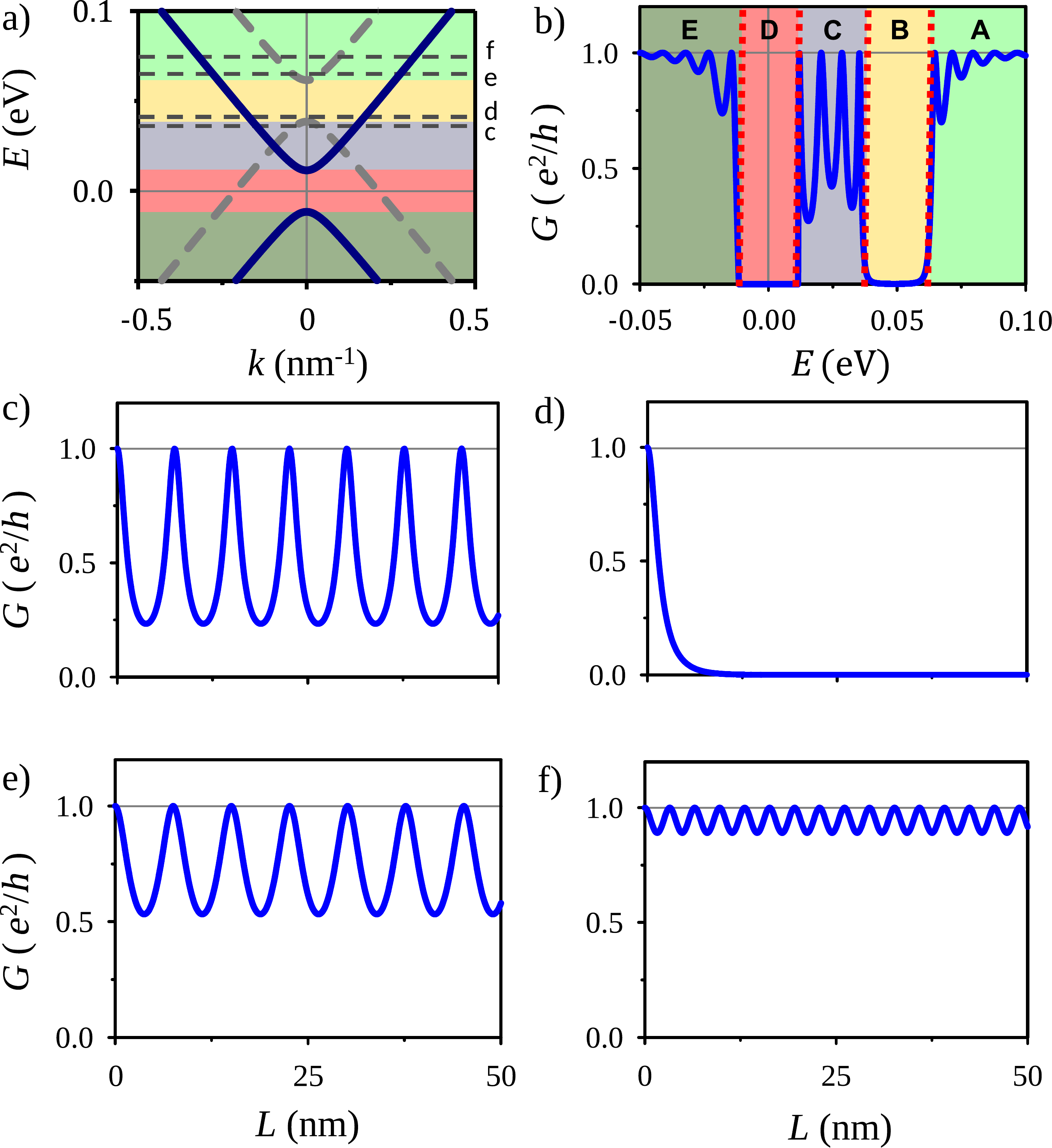}
	\caption{
		a) A pair of propagating bands ($j = 1/2$, $\nu = \pm 1$) of a cylindrical 3D~TI nanowire with radius $R = 10\,\textnormal{nm}$, surface-state Dirac velocity $v_\textsc{f} = 3.5 \times 10^5\,\textnormal{m/s}$, and a gated central section with length $L = 80\,\textnormal{nm}$. The spectrum of surface states in the leads and gated central section ($\Delta E = 50\,\textnormal{meV}$) are shown in solid blue lines and dashed gray lines, respectively.
		b) The conductance as a function of the energy of the 
		left-lead
		input mode  that is shown in a). The different transmission regimes, as discussed in section~\ref{Results_FP}, are indicated with different colors and labeled as in figure~\ref{F1}.
		c)-f) The conductance as a function of the gate length $L$ for the input-mode energies indicated by the horizontal dark-gray dashed lines in a) and labeled by the corresponding subfigure label. These are c) $E = 35\,\textnormal{meV}$, d) $E = 40\,\textnormal{meV}$, e) $E = 65\,\textnormal{meV}$, and f) $E = 75\,\textnormal{meV}$.
		Details on the model can be found in section~\ref{Methods_2D}.
	}
	\label{F2}
\end{figure}

We show the conductance in the different transmission regimes in figure~\ref{F2} as a function of the energy of the input mode
as well as the length of the central section. 
As inputs we consider all propagating
modes in the left lead
with a rightward propagation.  
As expected, a FP interference pattern appears whenever there are propagating modes in the central section. In general, the amplitude of the oscillation is largest in the energy windows where the wave number of the surface-state mode in the lead or the central section approaches zero, and vanishes for propagating modes with energy far above or below the Dirac point energy. The maximal amplitude is retrieved in the $npn$-junction configuration near the transition to regime B or D.

We also find that these interference patterns apply to 3D~TI nanowires with arbitrary cross sections, even though (generalized) angular momentum can no longer be considered as a conserved quantum number and the analytical solutions do not apply. Nonetheless, the results carry over well to 3D~TI nanowires with arbitrary cross sections (as long as the extension of the bulk is significantly larger than the penetration depth of the surface states) with the following substitution: $2 \pi R \rightarrow P$, with $P$ the perimeter of the cross section, which can be verified by comparing with the solutions of the 3D model that can be obtained numerically.

Fabricating small gates as suggested in figure \ref{F1}a
with, e.g., chemical etching is nontrivial
but feasible with today's advanced nanolithographic techniques.
Nonetheless,
in an experimental setup, it will be easier to vary the gate voltage of a single gate as compared to adjusting the gate length by depositing multiple gates on top of the TI nanowire. It is expected that the energy of the propagating modes in the central section will be shifted as a function of the gate (a more detailed analysis is presented in section~\ref{Results_characterization}). This will also modify the wave number difference between modes in the leads and in the central section, and can thus also induce a FP interference pattern.
Except near $k = 0$ or at higher energies where nonlinear corrections kick in, a linear dispersion is expected for the surface-state subbands and transmission oscillations can be expected as a function of the energy shift with a constant period of energy given by $\hbar v_\textsc{f} \pi / L$, with $L$ the length of the gated central section.

\begin{figure}[tb]
	\centering
	\includegraphics[width=0.7\linewidth]{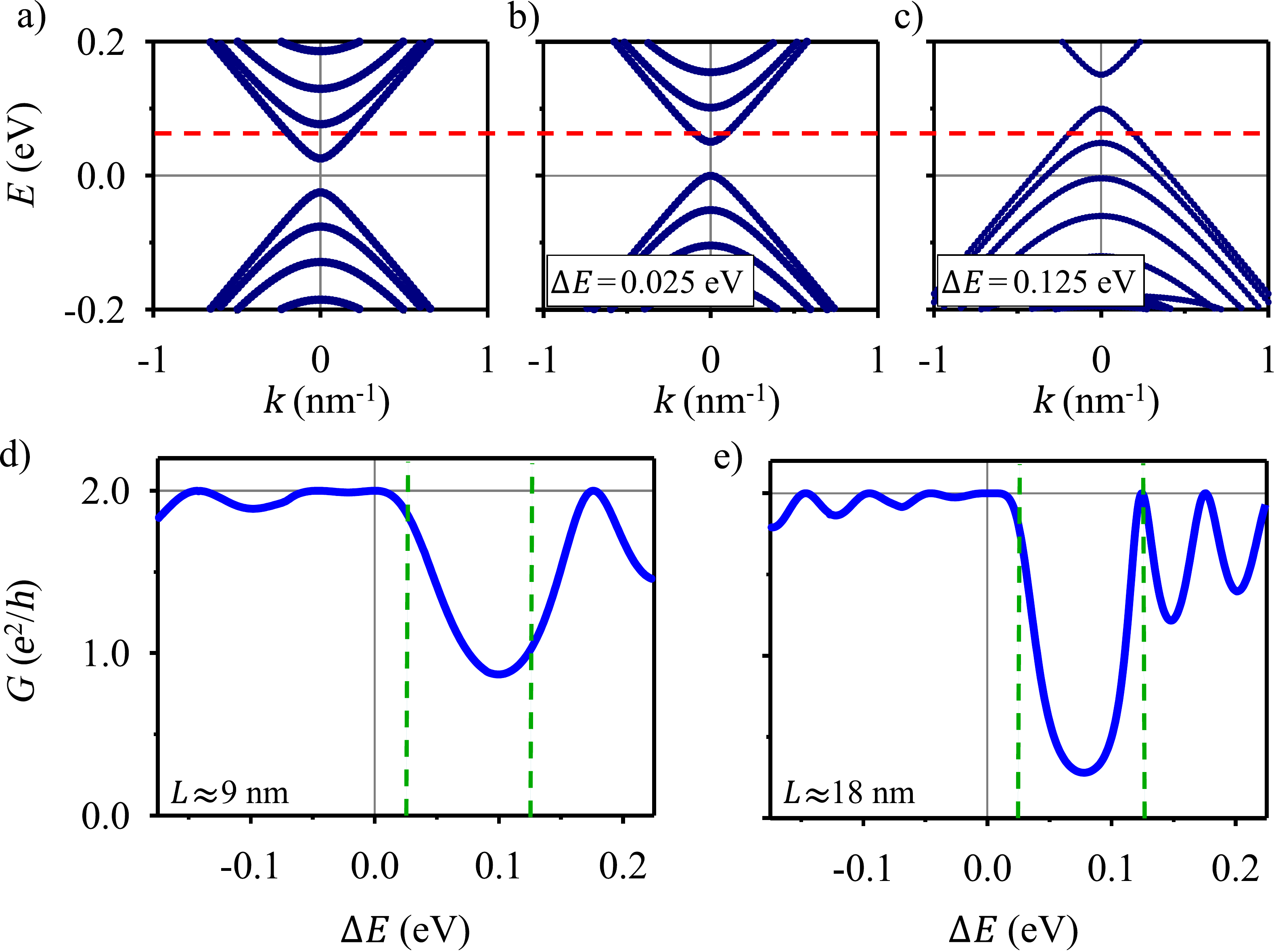}
	\caption{
		a)-c) Propagating bands of a) the leads and b),c) gated central section of a 3D~TI nanowire with a square-shaped cross section ($10 \times 10\,\textnormal{nm}^2$), considering the model of section~\ref{Methods_3D} and an energy shift of b) $\Delta E = 25\,\textnormal{meV}$ and c) $\Delta E = 125\,\textnormal{meV}$ in the central section with respect to the leads.
		d),e) The conductance as a function of the energy shift in the central section for central-section lengths d) $L \approx 9\,\textnormal{nm}$ and e) $L \approx 18\,\textnormal{nm}$.
		We consider input modes with energy equal to $E = 65\,\textnormal{meV}$ (indicated by the horizontal red dashed line).
		The values for the energy shift that are considered in b) and c) are indicated by vertical green dashed lines.
	}
	\label{F3}
\end{figure}

In figure~\ref{F3}, we can see the conductance of a noncylindrical 3D~TI nanowire as a function of the energy shift in the gated section for two different region lengths. When the energy shift is negative, the junction remains in the $nnn$-junction configuration (transmission regime A) for the input-mode energy under consideration and only conductance oscillations with very small amplitude are obtained. For an energy shift ranging between $\Delta E = 0.04\,\textnormal{eV}$ and $\Delta E = 0.09\,\textnormal{eV}$, the device is in the $n\mathrm{b}n$-junction configuration (transmission regime B) with evanescent modes in the gated central section. As a consequence, we can see a significantly suppressed conductance in both figures~\ref{F3}d and \ref{F3}e in this range of energy shifts. The conductance does not completely drop to zero because the gate length is not large enough to suppress tunneling through the central section. For an energy shift exceeding $0.09\,\textnormal{eV}$, we enter the $npn$-junction transmission regime (regime C) with pronounced oscillations of the conductance, as expected. We can see how these oscillations decrease again with increasing $\Delta E$. Also note how the oscillation period in terms of $\Delta E$ is different between figures~\ref{F3}d and \ref{F3}e due to different gate lengths being considered.

\subsubsection{Magnetic effects} \label{Results_magnetic_effects}
In this section, we consider the 3D~TI nanowire-based FP interferometer of figure~\ref{F1} in the presence of an external magnetic field that is aligned along the wire. We only consider the orbital effect on the surface-state charge carriers, as the Zeeman effect is typically negligible in comparison.
The orbital effect breaks the $l \leftrightarrow - l - 1$ degeneracy of the surface-state spectrum due to the flux dependency of $j$ [see (\ref{eq:2D_Hamiltonian}) and paragraph below]. While the FP interference for the $l$ and $-l-1$ modes is identical without an external magnetic field being present, their interference patterns deviate due to the flux, with the corresponding wave numbers in the central section splitting up: $k_\textsc{c} \rightarrow \{ k_\textsc{c}^+, k_\textsc{c}^- \}$ with $|k_\textsc{c}^+| > |k_\textsc{c}^-|$. For small field strengths, the wave number splitting is small and a beating pattern in the FP interference pattern emerges, with a periodicity in terms of central-section length equal to $L_\mathrm{p} = 2\pi / (|\Delta k_\textsc{c}^+| - |\Delta k_\textsc{c}^-|) = \pi / (|k_\textsc{c}^+| - |k_\textsc{c}^-|)$. The periodicity of this beating pattern is on a much larger length scale as compared to the single-mode conductance periodicity.

\begin{figure*}[tb]
	\centering
	\includegraphics[width=\linewidth]{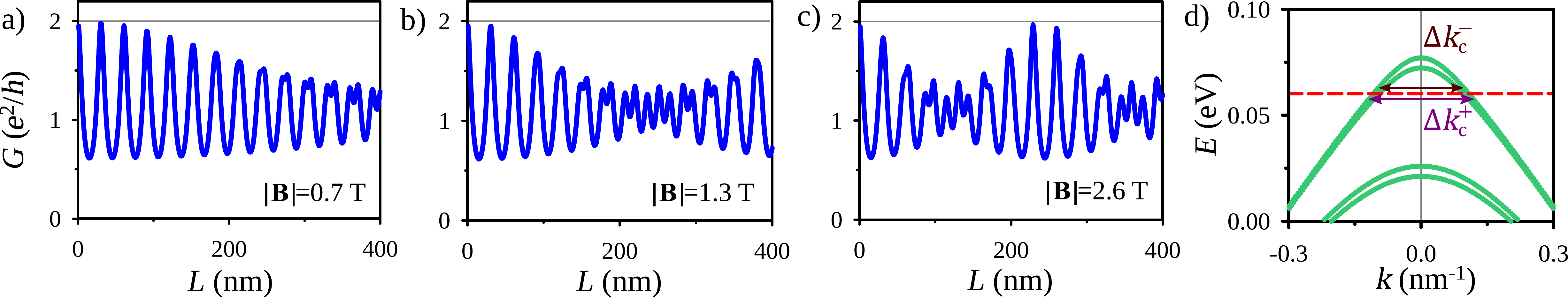}
	\caption{
		a)-c) The conductance of a 3D~TI nanowire with square-shaped cross section ($10 \times 10\,\textnormal{nm}^2$) as a function of the central-section length $L$ in the presence of an external magnetic field aligned with the wire, considering different field strengths and an input-mode energy of $E = 60\,\textnormal{meV}$, with an energy shift of $\Delta E = 100\,\textnormal{meV}$ in the gated central section.
		d) The energy spectrum of the propagating modes in the central section of the wire, with input-mode energy indicated by the horizontal red dashed line and aligned magnetic field with $|\mathbf{B}|=2.6\,\textnormal{T}$. The wave number differences between the forward- and backward-propagating modes, as introduced in section~\ref{Results_magnetic_effects}, are also indicated.
		Details on the model (parameters) can be found in section~\ref{Methods_3D}.
	}
	\label{F4}
\end{figure*}

In figure~\ref{F4}, the conductance is presented as a function of gate length for three different external magnetic fields strengths, considering a 3D~TI nanowire-based FP interferometer in the $npn$-junction configuration where the amplitude of the conductance oscillations is maximal. We can see single-mode FP oscillations with a periodicity $L_\mathrm{p} \sim 30\,\textnormal{nm}$, which are barely affected by the magnetic field, as well as a beating pattern with a periodicity of several hundreds of nm that is strongly dependent on the nanowire-piercing magnetic flux. While the explanation above is based on the surface-state model for cylindrical nanowires, the interpretation based on the wave number splitting also applies to 3D~TI nanowires with arbitrary cross sections (see figure~\ref{FA1} in \ref{A2} for the wave number splitting in the numerical approach where the wave numbers are generally complex values).

\begin{figure}[tb]
	\centering
	\includegraphics[width=0.45\linewidth]{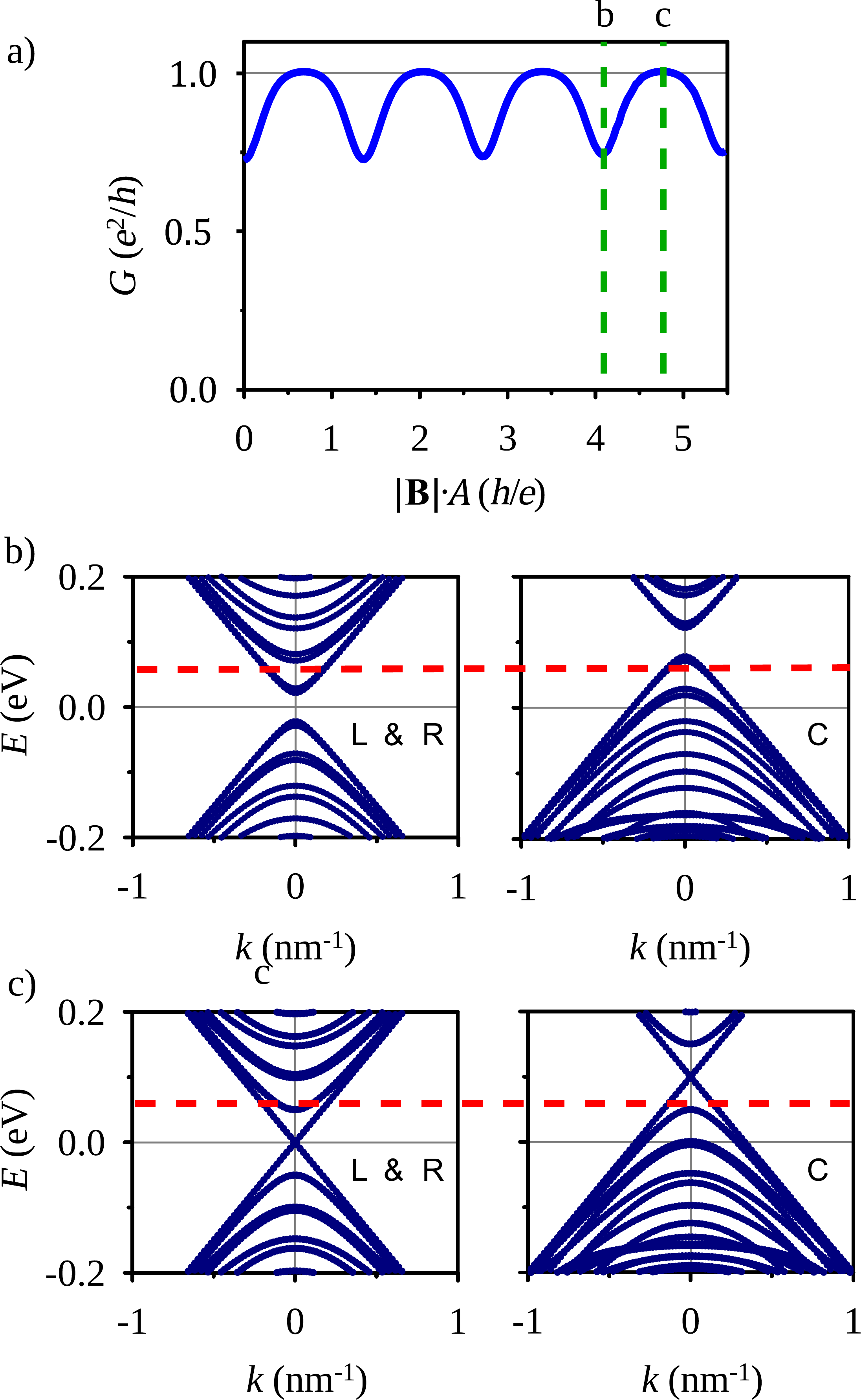}
	\caption{
		a) The conductance of a 3D~TI nanowire with square-shaped cross section ($10\times10\,\textnormal{nm}^2$) as a function of the magnetic flux piercing the cross section of the wire in units of the flux quantum.
		Vertical green dashed lines indicate the conductance for a given flux with the corresponding surface-state spectra in the gated section and leads in b) and c).
		b),c) The surface-state spectrum in (left) the leads and (right) gated central section with length $L = 20\,\textnormal{nm}$, considering a magnetic field strength equal to b) $|\mathbf{B}| A/(h/e) \approx 4$ and c) $|\mathbf{B}| A/(h/e) \approx 4.75$, where $A$ represents the cross-sectional surface area of the nanowire. The input-mode energy is equal to $E = 60\,\textnormal{meV}$ and an energy shift equal to $100\,\textnormal{meV}$ is considered in the gated central section. More details on the model (parameters) can be found in section~\ref{Methods_3D}.
		The input-mode energy is indicated with horizontal red dashed lines.
	}
	\label{F5}
\end{figure}

In the regime with strong external magnetic field, i.e., the flux quantum-piercing regime, we can expect the well-known flux quantum-periodic Aharonov-Bohm (AB) oscillations and the appearance of a single perfectly transmitting mode near the Dirac point (the regime where the only counterpropagating mode has opposite spin and backscattering is thus prohibited by time reversal symmetry).

In figure~\ref{F5}, we can see the flux quantum-periodic conductance oscillations up to a rescaling factor $\sim 1.35$ due to the finite extension of the surface states into the bulk of the wires, which reduces the flux that is effectively enclosed by the surface states. When the surface states effectively enclose a half-integer magnetic flux quantum, a gapless linear dispersion relation is recovered for one of the surface-state modes and the gated section becomes completely transparent for that mode. This effect is well known as Klein tunneling. In other words, transport regimes A-E, as specified in section~\ref{Methods_2D} and indicated in figures~\ref{F2}a and \ref{F2}b, do not apply to the gapless linear ($j = 0$) surface-state mode, as there is perfect transmission for all input energies (up to energies for which the description with linear dispersion is valid).

\subsubsection{Transverse asymmetry} \label{Results_transverse_asymmetry}
Now we consider the impact of transverse asymmetry of the gate effect on the FP interference pattern and the magnetic field dependence of the surface-state conductance. This effect will be present to some extent in a realistic device with a gate that does not completely wrap around the wire, e.g., a top-gated device~\cite{Ziegler2018}. We will proceed with a numerical analysis of transverse asymmetry by considering the 3D~TI continuum model of section~\ref{Methods_3D} with a transverse profile for the energy shift $\Delta E$ along the $z$-direction in the central section of the FP interferometer. We consider a nanowire with rectangular-shaped cross section, with dimensions $(L_y, L_z)$, and an asymmetry profile given by
\begin{equation} \label{E35}
	\Delta E(z) \equiv \Delta E +  \gamma (z - z_0) / L_z\, ,
\end{equation}
with $z_0$ the center of the wire along the $z$-direction and $\gamma$ an asymmetry parameter that characterizes the total transverse variation of the energy shift with $\Delta E(z) \in [\Delta E - \gamma/2, \Delta E + \gamma/2]$.
We summarize the impact on the FP conductance oscillations, as well as on the beating pattern and AB oscillations due to an external magnetic field, below.

\begin{figure}[tb]
	\centering
	\includegraphics[width=0.9\linewidth]{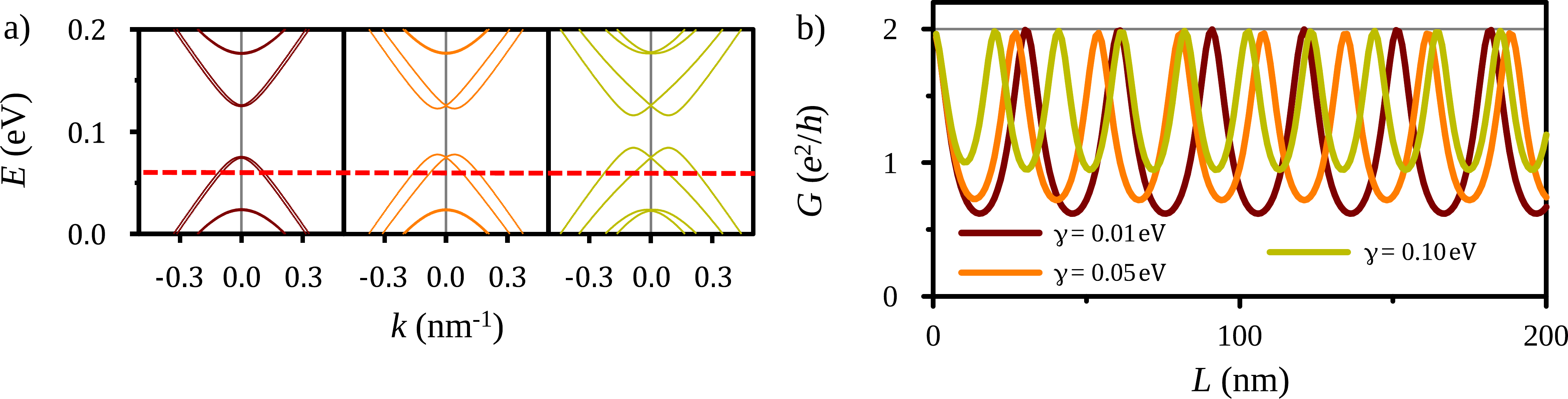}
	\caption{
		a) The energy spectrum of propagating bands in the gated section of a 3D~TI nanowire with square-shaped cross section ($10\times10\,\textnormal{nm}^2$) and a mean value of $\Delta E$ over the cross section equal to $\langle \Delta E \rangle = 0.1\,\textnormal{eV}$ in presence of different degrees of asymmetry. Using the parametrization for asymmetry of (\ref{E35}), we present the spectrum with $\gamma = 0.01\,\textnormal{eV}$, $\gamma = 0.05\,\textnormal{eV}$, and $\gamma = 0.1\,\textnormal{eV}$, from left to right, respectively.
		b) The conductance as a function of the central-section length for different strengths of the asymmetry parameter.
		Details on the model can be found in section~\ref{Methods_3D}.
	}
	\label{F6}
\end{figure}

\begin{figure}[tb]
	\centering
	\includegraphics[width=0.6\linewidth]{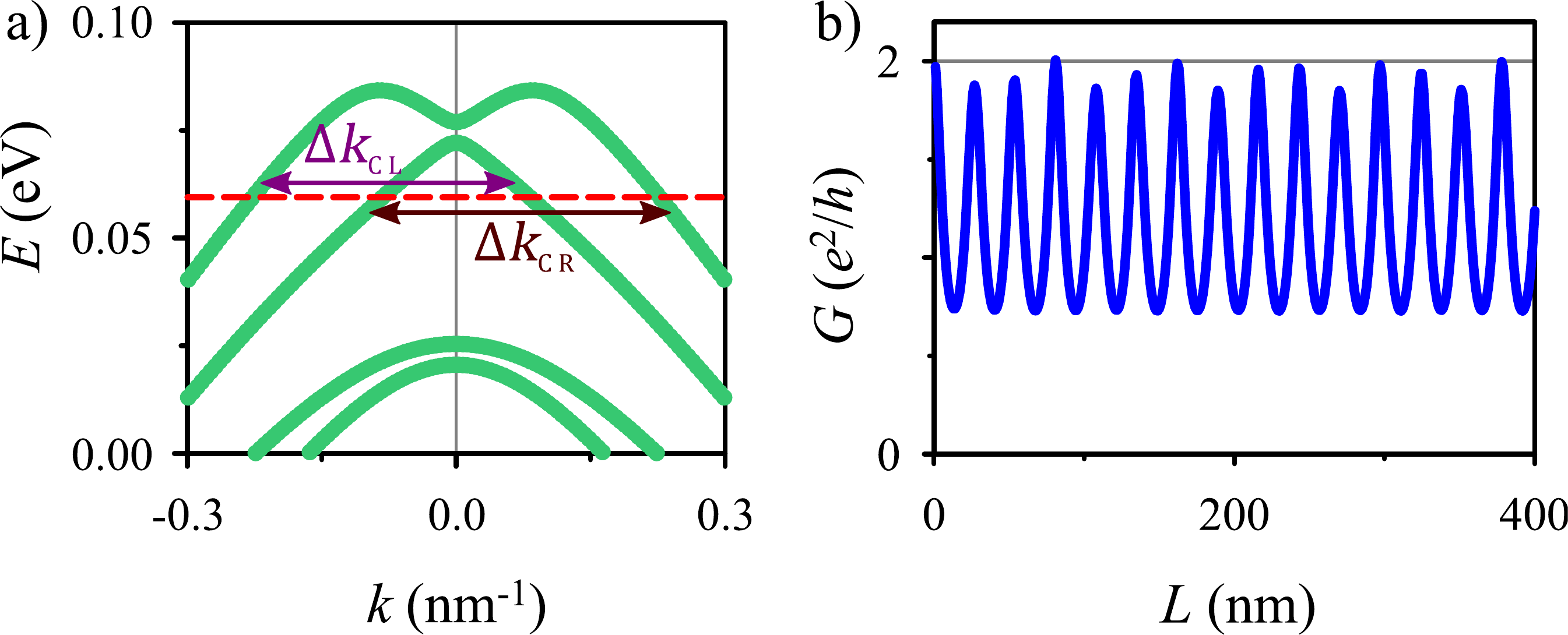}
	\caption{
		a) Propagating bands in the central section of a 3D~TI nanowire with square-shaped cross section ($10\times10\,\textnormal{nm}^2$) and a mean value of the energy shift over the cross section equal to $\langle \Delta E \rangle = 100\,\textnormal{meV}$, asymmetry parameter $\gamma = 0.1\,\textnormal{eV}$, and an external magnetic field aligned with te wire with field strength $|\mathbf{B}| \approx 2.6\,\textnormal{T}$.
		The wave number differences between the different forward- and backward-propagating modes in the central section, as explained in section~\ref{Results_transverse_asymmetry}, are also indicated.
		b) The conductance of the device described in a) as a function of the central-section length, considering an input-mode energy equal to $E = 60\,\textnormal{meV}$, as indicated by the horizontal red dashed line in a), and model parameters as specified in section~\ref{Methods_3D}.
	}
	\label{F7}
\end{figure}

\begin{figure}[tb]
	\centering
	\includegraphics[width=0.55\linewidth]{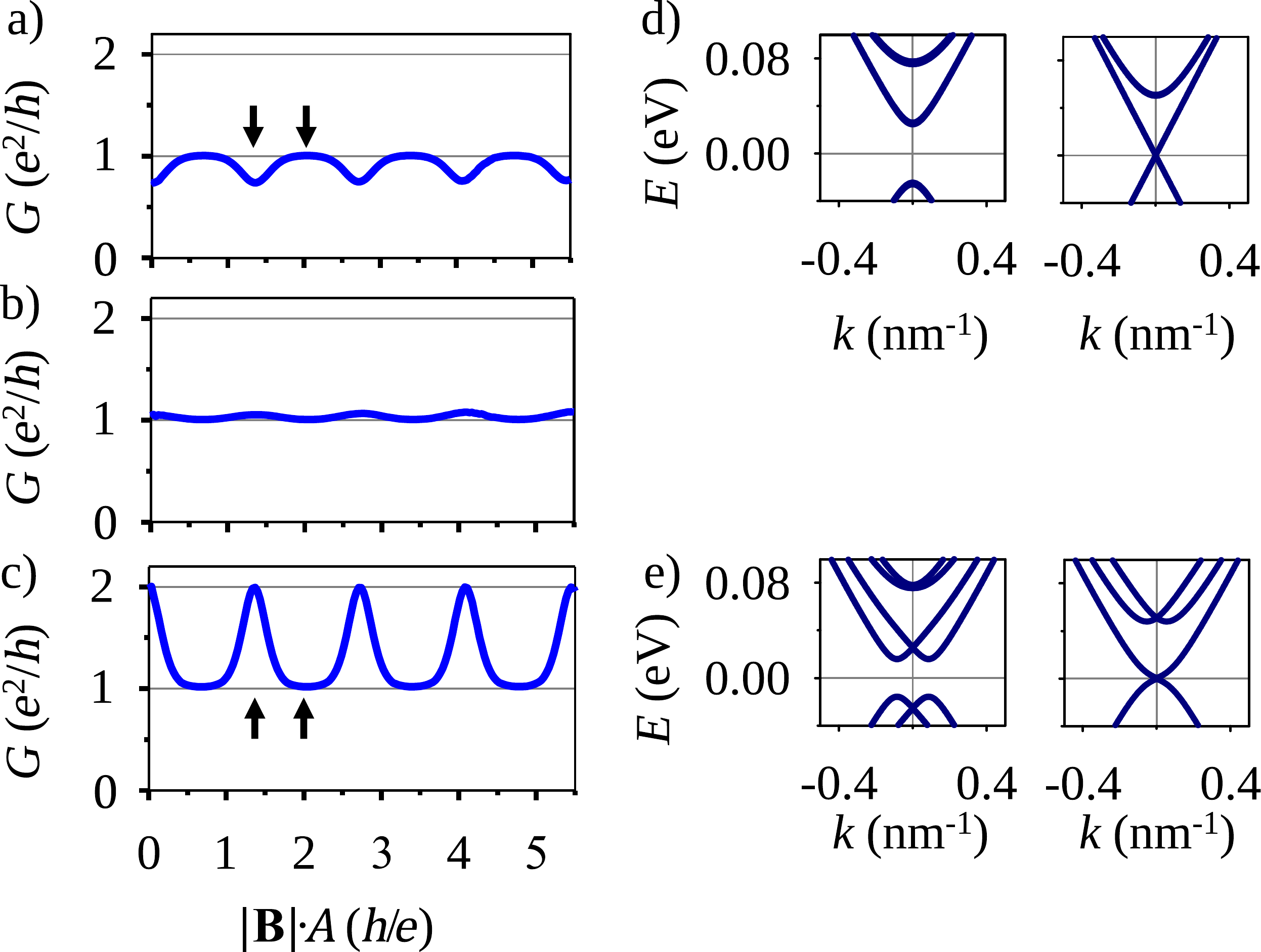}
	\caption{
		a)-c) The conductance of a 3D~TI nanowire with square-shaped cross section ($10\times10\,\textnormal{nm}^2$) as a function of the magnetic field with asymmetry parameter in the gated central section [see (\ref{E35})] equal to a) $\gamma = 0.01\,\textnormal{eV}$, b) $\gamma = 0.05\,\textnormal{eV}$, and c) $\gamma = 0.1\,\textnormal{eV}$, further assuming an input-mode energy $E = 60\,\textnormal{meV}$, a central-section length $L = 20\,\textnormal{nm}$, an energy shift in the gated central section of $100\,\textnormal{meV}$, and model parameters as specified in section~\ref{Methods_3D}.
		d),e) The energy spectrum of propagating bands in the gated central section of the 3D~TI nanowire device described in a) for asymmetry parameters d) $\gamma = 0.01\,\textnormal{eV}$ and e) $\gamma = 0.1\,\textnormal{eV}$ with a magnetic flux equal to (left) $|\mathbf{B}| A/(h/e) \approx 1.5$ and (right) $|\mathbf{B}| A/(h/e) \approx 2$ piercing the nanowire cross section.
		These flux values are indicated by black arrows in a) and c).
	}
	\label{F8}
\end{figure}

\begin{enumerate}[label=(\roman*)]
\item
\emph{Conductance oscillations without magnetic field} --
The asymmetry induces a splitting of the doubly-degenerate subbands in reciprocal space and an upward (downward) shift in energy for the hole (electron) bands, as shown in figure~\ref{F6}a, with a crossing at $k = 0$ (in the absence of an external magnetic field).
The FP conductance oscillations are preserved in the presence of asymmetry but, as shown in figure~\ref{F6}b, there is a decrease of the oscillation period (in terms of central-section length) and amplitude of the oscillations for increasing $\gamma$ due to the energy shift of the propagating modes and the corresponding increase of $|\Delta k_\textsc{c\,\textsc{l}}|$, $|\Delta k_\textsc{c\,\textsc{r}}|$, with $\Delta k_{\textsc{c}\,\textsc{l}}, \Delta k_{\textsc{c}\,\textsc{r}}$ the wave number differences between the forward- and backward-propagating modes of the subbands that have been shifted to the left and right, denoted by subscript $\textsc{l}$ and $\textsc{r}$, respectively (see figure~\ref{F6}a).
Note that $\Delta k_{\textsc{c}\,\textsc{l},\textsc{r}} \neq 2|k_{\textsc{c}\,\textsc{l},\textsc{r}}|$ in the presence of transverse asymmetry.

\item
\emph{Beating pattern for small magnetic fields} --
The beating pattern of the conductance oscillations with a large period as a function of central-section length, originating from the small splitting of $k_\textsc{c} \rightarrow \{k_\textsc{c}^+, k_\textsc{c}^-\}$ and resulting difference of the FP conductance oscillation periods, is quenched by transverse asymmetry (see figure~\ref{F7}). Due to the splitting of the degenerate subbands in reciprocal space, a small magnetic field does not induce a significant difference between $\Delta k_{\textsc{c}\,\textsc{l}}$ and $\Delta k_{\textsc{c}\,\textsc{r}}$, as is the case for the transverse-symmetric system. In other words, the wave number differences remain comparable, $\Delta k_{\textsc{c}\,\textsc{l}} \approx \Delta k_{\textsc{c}\,\textsc{r}}$ in the presence of a small magnetic field (see figure~\ref{F7}a and compare to $\Delta k_\textsc{c}^-$ and $\Delta k_\textsc{c}^+$ in figure~\ref{F4}d).
The two oppositely shifted subbands do not remain completely decoupled in the presence of an external magnetic field, however, so there is some intersubband coupling (with a different periodicity) which is responsible for the (minor) differences between figure~\ref{F6}b and figure~\ref{F7}b, and the gap-opening at $k = 0$ where the two subbands cross.

\item
\emph{Aharonov-Bohm oscillations for large magnetic fields} --
We find that the above-discussed Klein tunneling regime with $G = e^2/h$ when the surface states enclose a half-integer magnetic flux quantum are preserved when introducing transverse asymmetry in the gated section of the nanowire. Away from this condition, the conductance remains sensitive to the central-section length and energy of the propagating mode, as is the case for the transverse-symmetric setup, and is sensitive to the specific value of the asymmetry parameter as well, as can be seen in figure~\ref{F8}.
\end{enumerate}

Interestingly, the surface-state spectrum of a 3D~TI nanowire with transverse asymmetry and an aligned magnetic field (see figure~\ref{F7}a) is qualitatively similar to the spectrum of Rashba nanowires in the presence of an external magnetic field, with the appearance of an anticrossing (chiral gap) between two bands that are shifted in reciprocal space. The 1D model Hamiltonian $H_\textsc{r}(k)$ for Rashba nanowires is given by,
\begin{equation}
	H_\textsc{r}(k) = \hbar^2 k^2 / (2 m_e^\ast) + (\alpha/\hbar) \sigma_y k + E_\textsc{z} \, \sigma_z\, .
\end{equation}
Spin-orbit coupling (SOC) is represented by the second term and the Zeeman effect by the third term. Note that the kinetic term (the first term on the right-hand side) is quadratic rather than linear as is the case for the 3D~TI surface states. SOC splits the spin-degenerate parabolic bands in reciprocal space, whereas the Zeeman effect opens up a gap where the two subbands would cross, i.e., an anticrossing is obtained. Peculiarly, a similar band structure is obtained for the lowest-energy subbands of a 3D~TI nanoribbon when including transverse asymmetry of the Fermi level energy. We can draw the following analogy between the parameters of the Rashba model on the one hand, and the 3D~TI nanowire surface-state model on the other hand:
\numparts
\begin{eqnarray}
	\hbar^2 k^2 / (2 m_e^\ast) & \leftrightarrow &(\hbar v_\textsc{f} P)/(2 \pi)\, , \\
	(\alpha/\hbar) & \leftrightarrow & \gamma\, , \\
	E_\textsc{z} & \leftrightarrow & (\Phi\,\,{\rm mod}\,\,\Phi_0)\, .
\end{eqnarray}
\endnumparts
Note that, while $\mathbf{\sigma}$ typically acts on the spin-$\pm 1/2$ subspace of semiconductor-nanowire bulk modes in the Rashba model, it acts on the $\pm j$ transverse-mode subspace of 3D~TI nanowire surface states on the other side of the analogy.
Further note that this chiral gap is commonly considered for the realization of a topological spinless $p$-wave superconductor in semiconductor nanowires with strong SOC~\cite{Lutchyn2010, Oreg2010}.
In proximitized 3D~TI nanowires, however, the conditions are rather different and such a chiral gap induced by transverse asymmetry is not necessarily required~\cite{Cook2011, Cook2012, deJuan2019}.
Nonetheless, it is a peculiar analogy as spin-orbit coupling is replaced by transverse asymmetry of the gate effect and corresponding energy shift, the Zeeman effect by the orbital effect of an external magnetic field, and the spin-$\pm$1/2 subspace by different transverse modes of the spin-momentum locked surface-state Dirac cone spectrum of a 3D~TI nanowire.
And it could be exploited to overcome the requirement of a vortex in the superconducting pairing potential in proximitized 3D~TI nanowires~\cite{Legg2021}.

\subsection{Experimental characterization} \label{Results_characterization}
In this section, we consider the above-mentioned transmission properties and conductance oscillations for characterizing 3D~TI nanoribbon surface states. A detailed characterization of the energy shift in the gated section with respect to the Dirac point energy could be useful for tuning the 3D~TI nanoribbon in the ideal regime for realizing proximity-induced topological superconductivity, i.e., the regime with an odd number of surface-state subband crossings, and realizing the intensely sought-after Majorana bound states~\cite{Cook2011}.

\begin{figure}[tb]
	\centering
	\includegraphics[width=0.7\linewidth]{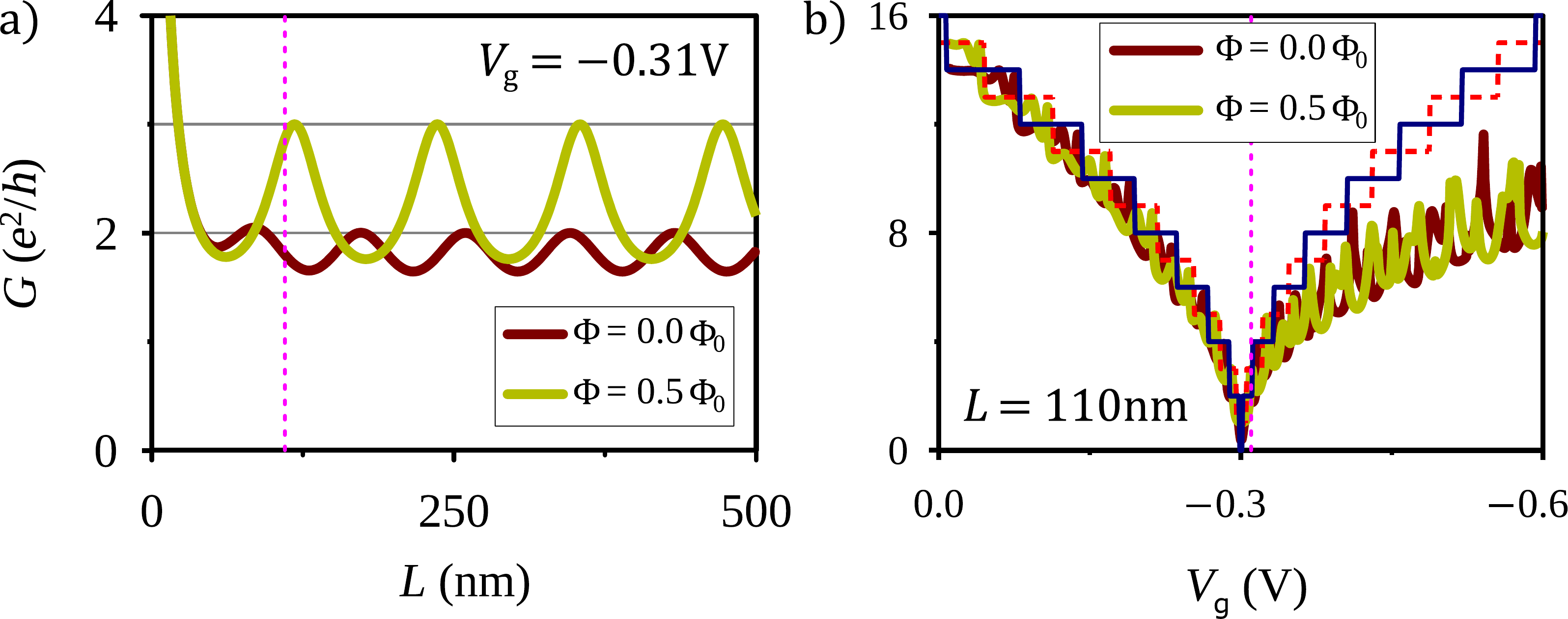}
	\caption{
		a),b) The conductance of 3D~TI nanoribbon surface states across a gated central section as a function of a) the length of this section (with $V_\mathrm{g} = -0.31\,\textnormal{V}$) and of b) the gate voltage (with $L = 110\,\textnormal{nm}$), considering (solid red lines) zero magnetic field and (solid green lines) a magnetic field that pierces the nanoribbon cross section with half a magnetic flux quantum.
		The solid (dashed) steplike lines indicate the number of modes in the central section for zero magnetic field  (the $\Phi_0/2$-pierced regimes).
		The dotted purple line indicates the considered gate voltage or length in the other subfigure.
	}
	\label{F9}
\end{figure}

Based on reasonable assumptions regarding the experimental feasibility, we consider a gated 3D~TI nanoribbon with rectangular cross section and with the following device and surface-state characteristics: $E_\textsc{f} = 50\,\textnormal{meV}$, $v_\textsc{f} = 3.5 \times 10^5\,\textnormal{m/s}$, $W = 100\,\textnormal{nm}$, $H = 10\,\textnormal{nm}$, $P = 220\,\textnormal{nm}$, $C_\mathrm{eff} = 2 \times 10^{-3}\,\textnormal{F/m}^2$, with $W, H, P$ respectively the width, height, and perimeter ($P = 2W + 2H$) of the nanoribbon cross section, and $C_\mathrm{eff}$ the effective capacitance of the gated section per unit area~\cite{Lanius2016, Ziegler2018, Weyrich18, Koelzer2020, Rosenbach2020, Rosenbach2021}.
The surface-state energy spectrum in the central section is shifted as a function of the gate voltage $V_\mathrm{g}$. As before, we denote the energy shift by $\Delta E$ such that we obtain
\begin{equation}
	s (E_\textsc{f} + \Delta E)^2 / (\hbar v_\textsc{f})^2 = E_\textsc{f}^2 / (\hbar v_\textsc{f})^2 + 4 \pi C_\mathrm{eff} V_\mathrm{g}/e\, ,
\end{equation}
yielding
\begin{equation}
	\Delta E = -E_\textsc{f} + s \sqrt{\left| E_\textsc{f}^2 + 4 \pi (\hbar v_\textsc{f})^2 C_{\mathrm{eff}} V_\mathrm{g}/e \right|}\, ,
\end{equation}
with $s \equiv \mathrm{sgn} (E_\textsc{f}^2 / (\hbar v_\textsc{f})^2 + 4 \pi C_\mathrm{eff} V_\mathrm{g}/e)$ and having made use of the following relation between the Fermi wave vector $k_\textsc{f} = |E_\textsc{f}| / \hbar v_\textsc{f}$ (which gets shifted by $\Delta E/(\hbar v_\textsc{f})$) and the 2D charge carrier density $n$ (which gets shifted by $C_{\mathrm{eff}} V_\mathrm{g}/e$) for a spin-nondegenerate Dirac cone: $k_\textsc{f}^2 = 4 \pi n$.
Note that we approximate the density of states of the subband-quantized spectrum by the 2D density of states of a 2D surface-state Dirac cone, as $E_\textsc{f} \gg 2 \pi \hbar v_\textsc{f}/P \approx 6.6\,\textnormal{meV}$, with the latter being the energy spacing between surface-state subbands at $k = 0$. The gate voltage $V_\mathrm{g}^{(\textsc{d})}$ that tunes the surface-state spectrum in the central section to the Dirac point is given by,
\begin{equation}
	V_\mathrm{g}^{(\textsc{d})} = -e E_\textsc{f}^2/[4 \pi (\hbar v_\textsc{f})^2 C_\mathrm{eff}] \approx -0.3\,\mathrm{V}\, .
\end{equation}
Ideally, the gate length must be chosen such that multiple FP oscillations of the lowest-energy surface-state mode can be realized as a function of the gate voltage before an additional surface-state mode has access to a propagating mode in the central section. The periodicity of the FP oscillations can be obtained from the following condition: $\Delta(2 k_\textsc{c} L) = 2 \pi$. Approximating the spectrum by $E(k) = \hbar v_\textsc{f} |k|$, we obtain $k_\textsc{c} \approx |E_\textsc{f} + \Delta E| / (\hbar v_\textsc{f}) = \sqrt{\left| E_\textsc{f}^2 / (\hbar v_\textsc{f})^2 + 4 \pi C_\mathrm{eff} V_\mathrm{g}/e \right|}$. The oscillations can be resolved as a function of the gate length or, alternatively, as a function of the gate voltage (see figure~\ref{F9}), with the latter being more suitable for an experimental setup, as it can easily be varied continuously on a single sample.
The oscillation period as a function of gate length, denoted by $\Delta L$, is given by
\begin{equation}
	\eqalign{\Delta L &= \pi / k_\textsc{c} = \pi \hbar v_\textsc{f} / (E_\textsc{f} + \Delta E) \\
		&\approx 14.5\,\mathrm{nm} \times \frac{1}{1 + \Delta E/E_\textsc{f}}\, .}
\end{equation}
Note that $\Delta L \rightarrow +\infty$ when the Dirac point aligns with the Fermi level ($E_\textsc{f} + \Delta E \rightarrow 0$), as the wave number difference vanishes (ignoring deviations due to subband quantization).

\begin{figure*}[tb]
	\centering
	\includegraphics[width=0.85\linewidth]{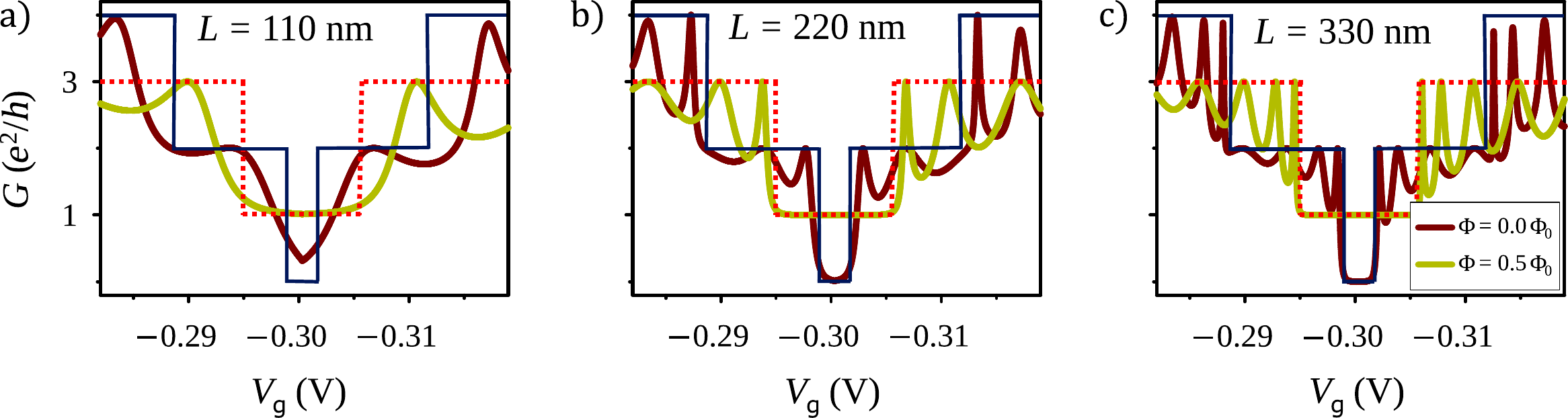}
	\caption{
		a)-c) The conductance of 3D~TI nanoribbon surface states across a gated central section as a function of the gate voltage for central-section lengths a) $L = 110\,\textnormal{nm}$, b) $L = 220\,\textnormal{nm}$, and c) $L = 330\,\textnormal{nm}$. The results are obtained for (red solid lines) zero magnetic field as well as for (green solid lines) a magnetic field that pierces the nanoribbon cross section with half a magnetic flux quantum. The solid blue (red dashed) lines indicate the number of propagating modes in the central section for zero flux (half a flux quantum) piercing the cross section.
	}
	\label{F10}
\end{figure*}

The oscillation period as a function of gate voltage, denoted by $\Delta V_\mathrm{g}$, depends on the Fermi level and thereby also on the reference value that is considered for the applied gate voltage. We consider the oscillation period near the Dirac point with $\left| E_\textsc{f}^2/(\hbar v_\textsc{f})^2 + 4 \pi C_\mathrm{eff} V_\mathrm{g}/e \right| = 0$, yielding
\begin{equation}
	\Delta V_\mathrm{g} = \left| e (\pi/L)^2/(4 \pi C_\mathrm{eff}) \right|\, .
\end{equation}
On the other hand, the gate voltage shift $\Delta V_\mathrm{g}^{(\mathrm{sb})}$ that is required to induce an additional subband crossing near the Dirac point is given by:
\begin{equation}
	\Delta V_\mathrm{g}^{(\mathrm{sb})} = \left| e (2 \pi / P)^2/(4 \pi C_\mathrm{eff}) \right|\, .
\end{equation}
By comparing the two, we can see that at least $L > P/2$ is required, and preferably $L > P$, to resolve FP oscillations as a function of gate voltage without changing the number of propagating modes in the central section. This requirement is confirmed by the transmission spectra as a function of gate voltage in figure~\ref{F10}. The gate voltage window is chosen such that the Dirac point is close to the Fermi level energy in the gated section of the wire. It can clearly be seen that additional FP conductance oscillations appear as a function of the gate voltage when $L > P/2$ and the number of FP oscillations in a gate voltage window with constant number of propagating modes agrees well with $2 L/P$. Furthermore, the half-integer flux-pierced regime shows an extended plateau with $G = e^2/h$ that is not affected by the gate, being a hallmark signature of Klein tunneling.

Note that the FP oscillations and the Klein tunneling offers more direct signatures of phase-coherent transport of 3D~TI surface states as compared to the AB-type magnetoconductance oscillations. AB oscillations are related to the subband spacing due to quantization of the momentum along the perimeter of the wire, but not directly to the dispersion relation along the transport direction. The signatures discussed can be used to map out the 3D~TI nanoribbon surface states in momentum space as well as pinpoint the Dirac point and characterize the precise number of 3D~TI surface-state band crossings in the gated section. While varying the gate length offers the most straightforward method to resolve FP oscillations, varying the gate voltage provides a suitable alternative due to the linear dispersion relation of 3D~TI surfaces states. For the latter, the FP resonances need to be disentangled from oscillations due to the variation of the number of propagating modes in the central section. For this, an appropriate gate length needs to be considered, with $L > P$ being ideal for obtaining several FP oscillations as a function of the gate voltage without changing the number of propagating modes in the central section.

Further note that the most pronounced FP-type conductance oscillations are obtained when the Fermi level energy in the central section is on the opposite side of the Dirac point as compared to the leads, i.e., an $npn$-junction in our example, and that the average conductance is significantly lower than for the $nnn$-junction configuration.

\section{Discussion \& Conclusion} \label{Discussion_Conclusion}
Before concluding, we discuss the findings and provide some additional remarks.
One important remark with respect to our device modeling approach is that we assume sharp steplike boundaries for the gated section. The question immediately arises whether this assumption is justified, especially when the surface-state spectrum is tuned near the Dirac point and, hence, few surface-state charge carriers are available to screen the gate effect outside the central section. In this regard, a 3D~TI sample with noninsulating bulk could be beneficial. In the Bi${}_2$Se${}_3$ family of 3D~TI materials, the bulk Fermi level typically crosses the valence or conduction band, and bulk charge carrier densities up to $\sim 10^{19}\,\textnormal{cm}^{-3}$ and above are no exception. Therefore, it is expected that the bulk charge carriers will efficiently screen the electric field outside of the central section and a fast decay (in the several $\si{\angstrom}$ to nm range) of the energy shift of the surface-state spectrum is expected along the wire direction.

For the experimental verification, it is also important to get into a transmission regime where the propagating surface-state modes in the central section do not suffer from elastic scattering. Due to the limited availability of states and spin-momentum locking, 3D~TI surface states are highly suitable to get into a quasiballisic transport regime over relatively long distances~\cite{Dufouleur2018}. Even in disordered 3D~TI nanowires with inter-defect distances in the few-nm range, a mean free path up to hundreds of nm can be achieved~\cite{Dufouleur2013, Cho2015}.

In conclusion, we have analyzed the conductance of a 3D~TI nanowire with a gated section where the surface-state spectrum can be shifted. This setup allows for Fabry-P\'erot interferometry of the surface-states, which reveals their phase-coherent transport properties and dispersion relation through conductance oscillations, which are most pronounced when the spectrum in the central section and leads is opposite with respect to the Dirac point.
Furthermore, the system can be driven into a Klein tunneling regime with a robust quantized conductance plateau with perfect transmission of the gapless surface-state subband by applying an external magnetic field along the wire that pierces the cross section with a half-integer flux quantum.
These signatures are robust for different cross sectional shapes and against transverse asymmetry of the gate effect, and they can be exploited for a detailed (magneto)transport-based characterization of 3D~TI surface states, which is promising for the realization of 3D~TI nanowire-based quantum devices.

\ack
L.S.\ acknowledges support from MINECO (Spain) Grant No. MAT2017-82639 and MINECO/AEI/FEDER Maria de Maeztu Program for units of Excellence MDM2017-0711.
K.M.\ acknowledges financial support from the Bavarian Ministry of Economic Affairs, Regional Development and Energy.
\newline

\bibliographystyle{iopart-num-mod}
\bibliography{imec3_rev}

\break
\appendix
\renewcommand\thefigure{A\arabic{figure}}
\setcounter{figure}{0}
\section{Details on 2D surface-state model} \label{A1}
The spinor components of the eigenstates presented in (\ref{eq:2D_eigenstates_A})-(\ref{eq:2D_eigenstates_B}) are given by
\begin{equation}
	\left( \matrix{
		g_\nu(j, k) \cr
		h_\nu(j, k) \cr
	} \right) = \left\{
	\matrix{ \left( \matrix{
			\sin\gamma_\nu(j, k)\cr
			\mathrm{i} \cos\gamma_\nu(j, k) \cr
		} \right) & (\nu j \geq 0) \cr
		\left( \matrix{
			\cos\gamma_\nu(j, k) \cr
			\mathrm{i} \sin\gamma_\nu(j, k) \cr
		} \right) & (\nu j \leq 0)
	}\right.\, ,
\end{equation}
when $|E| \geq |j/R|$, where
\begin{equation}
	\eqalign{\gamma_\nu(j, k) &\equiv \arctan\left[ \nu kR / \left(|j| + \sqrt{k^2 R^2 + j^2} \right) \right] \\
		&\hphantom{=} \in[-\pi/4, \pi/4]\, .}
\end{equation}
On the other hand when $|E| < |j/R|$ the solutions are evanescent modes which can be obtained from the plane-wave solutions by considering the substitution $k \rightarrow \mathrm{i} q$ ($-|j/R| \leq q \leq |j/R|$) yielding
\begin{equation}
	\left( \matrix{
		g_\nu(j, q) \cr
		h_\nu(j, q) \cr
	} \right) = \left\{\matrix{
		\left( \matrix{
			\sin\gamma_\nu(j, q) \cr
			\cos\gamma_\nu(j, q) \cr
		} \right) & (\nu j\geq 0) \cr
		\left( \matrix{
			-\cos\gamma_\nu(j, q) \cr
			\sin\gamma_\nu(j, q) \cr
		} \right) & (\nu j \leq 0)
	}\right.\, ,
\end{equation}
where
\begin{equation}
	\eqalign{\gamma_\nu(j, q) &\equiv \arctan\left[\left(|j| - \sqrt{j^2 - q^2 R^2}\right)/(\nu qR)\right] \\
		&\hphantom{=} \in [-\pi/4,\pi/4]\, ,}
\end{equation}
and with an energy expression $E_\nu(j, q) = \nu \hbar v_\textsc{f} \sqrt{j^2/R^2 - q^2}$.

The following solutions are obtained in the setup of the scattering problem:
\begin{equation}
	\Psi(x, \phi) = e^{\mathrm{i} k x + \mathrm{i} l \phi} \, \Phi_\nu(l, k) + r \, e^{-\mathrm{i} k x + \mathrm{i} l \phi} \, \Phi_\nu(l, -k)
\end{equation} 
when $(x \leq -L/2)$,
\begin{equation}
	\eqalign{\Psi(x, \phi) &= B_\textsc{R} \, e^{\mathrm{i} k' x - \mathrm{i} l \phi} \, \Phi_{\nu'}(l, k') \\
		&\hphantom{=} + B_\textsc{L} \, e^{-\mathrm{i} k' x - \mathrm{i} l \phi} \, \Phi_{\nu'}(l, -k')}
\end{equation}
when $(-L/2 \leq x \leq L/2,\,\textnormal{prop.\ modes})$,
\begin{equation}
	\eqalign{\Psi(x, \phi) &= C_\textsc{L} \, e^{-|q|(x + L/2) - \mathrm{i} l \phi} \, \Phi_{\nu'}(l, |q|) \\
		&\hphantom{=} + C_\textsc{R} \, e^{|q|(x - L/2) - \mathrm{i} l \phi} \, \Phi_{\nu'}(l,-|q|)}
\end{equation}
when $(-L/2 \leq x \leq L/2,\,\textnormal{evan.\ modes})$, and
\begin{equation}
	\Psi(x, \phi) = t \, e^{\mathrm{i} k x - \mathrm{i} l \phi} \, \Phi_\nu(l, k)
\end{equation}
when $(L/2 \leq x)$, where we need to consider different regimes in the gated section. For example, if $|E - \Delta E| < |\hbar v_\textsc{f} j/R|$, i.e., when $\mathrm{sgn}(U) (\Delta\epsilon^2 - \Delta\epsilon \, j/R)^{1/2} < k < \mathrm{sgn}(\Delta E) (\Delta\epsilon^2 + \Delta\epsilon \, j/R)^{1/2}$ with $\Delta\epsilon \equiv \Delta E / (\hbar v_\textsc{f})$, the solutions in the this region are evanescent modes instead of plane waves and $k' = \pm [k^2 + \Delta\epsilon^2 - 2 \, \mathrm{sgn}(k) \Delta\epsilon \sqrt{k^2 + j^2/R^2}]^{1/2}$ becomes imaginary.

We present the analytical expressions for the transmission coefficient $t$ for two transmission regimes as an example:
\begin{itemize}
	\item[(A)] $|E - \Delta E| \geq |\hbar v_\textsc{f} j/R|$ and $\nu = \nu'$:
	\begin{equation}
		t = \frac{2 \sin (2 \gamma_\textsc{l}) \sin (2 \gamma_\textsc{c}) e^{\mathrm{i} (k_\textsc{c} - k_\textsc{l}) L}}{1 - \cos (2 \gamma_\textsc{c} + 2 \gamma_\textsc{l}) - 2 \, e^{2 \mathrm{i} k_\textsc{c} L} \sin^2(\gamma_\textsc{c} - \gamma_\textsc{l})}\, .
	\end{equation}
	\item[(C)] $|E - \Delta E| \geq |\hbar v_\textsc{f}j/R|$ and $\nu = -\nu'$:
	\begin{equation}
		t = \frac{2 \sin (2 \gamma_\textsc{l}) \sin (2 \gamma_\textsc{c}) e^{\mathrm{i} (k_\textsc{c} - k_\textsc{l}) L}}{1 + \cos (2 \gamma_\textsc{c} - 2 \gamma_\textsc{l}) - 2 \, e^{2 \mathrm{i} k_\textsc{c} L} \cos^2(\gamma_\textsc{c} + \gamma_\textsc{l})}\, .
	\end{equation}
\end{itemize}
If $j = 0$, however, the gated section of the nanowire becomes transparent ($|t|^2 = 1$) due to perfect Klein tunneling, as expected for gapless surface-state bands with a linear dispersion relation (i.e., massless Dirac fermion solutions).

\section{Complex band structure} \label{A2}
The nanowire description based on the 3D~TI continuum model requires, as an initial step, the complex band structure $\{k, \phi_k\}$ in each homogeneous piece of the wire, with $k$ a complex-valued wave number and $\phi_k$ the corresponding wave function.
Here, we derive the non-Hermitian eigenvalue equation for the complex band structure. The approach amounts to introducing an extra pseudospin component and transforming the usual $E$-eigenvalue problem into a $k$-eigenvalue problem.

It is useful to rewrite the Hamiltonian of (\ref{E1}) with a given wave number $k_x \equiv k$ as
\begin{equation}
	{H}\equiv h_0 + h_x\, ,
	\label{eq10}
\end{equation}
where $h_0$ is the transverse Hamiltonian not depending on $k$ while the $k$ dependence is all contained in $h_x$, the longitudinal Hamiltonian,
\begin{eqnarray}
	\eqalign{h_0 &= 
	(C_0 - C_\perp\, k_y^2 - C_\parallel\, k_z^2) \\
	&\hphantom{=} +
	\tau_z\,
	(M_0 - M_\perp\, k_y^2-M_\parallel\, k_z^2) \\
	&\hphantom{=} +
	\tau_x\,
	(A_\perp\, \sigma_y k_y + A_\parallel\, \sigma_z k_z) \, ,}
	\\
	h_x =
	- (C_\perp+M_\perp \tau_z)\, k^2 + A_\perp \tau_x\sigma_x\, k \, .
\end{eqnarray}
Next, we also define an extra component $g = 1,2$ of the transverse wave function, in addition to the $\sigma,\tau = 1,2$ components. Simplifying the notation of wave function components $\phi_{\sigma\tau}(yz) \equiv \phi_k(yz,\sigma\tau)$, the generalization reads
\begin{equation}
	\phi_{\sigma\tau g}(yz)=\left\{
	\begin{array}{ll}
		\phi_{\sigma\tau}(yz) & {\rm if}\; g=1\\
		\rule{0cm}{0.6cm}
		\displaystyle\frac{C_\perp+M_\perp s_\tau}{E_U L_U}\,k\, \phi_{\sigma\tau}(yz) & {\rm if}\; g=2
	\end{array}
	\right.\, ,
\end{equation} 
where $s_{1,2} = \pm$ (for $\tau = 1,2$) and $E_U = 1 \,\textnormal{eV}$, $L_U = 1 \,\si{\angstrom}$ constants of energy and length, respectively, which can be chosen arbitrarily.
Notice that the new component $g=2$ is proportional to $k$. It allows to absorb the term of the Hamiltonian that is linear in $k$ into the wave function, while the quadratic term becomes linear in the enlarged space. The $k$-eigenvalue equation in this enlarged space reads

\begin{equation}
		\left\{
		\begin{array}{rcl}
			\displaystyle\frac{E_U L_U}{C_\perp + M_\perp s_\tau}\,
			\phi_{\sigma\tau 2}
			&=& 
			k\, 
			\phi_{\sigma\tau 1}\; ,\\
			\rule{0cm}{0.7cm}
			\displaystyle
			\frac{1}{E_U L_U}
			\sum_{\sigma'\tau'}
				\langle \sigma\tau
				|\,h_0\,|
				\sigma'\tau'\rangle\,
				\phi_{\sigma'\tau' 1}
				&+& \frac{E}{E_U L_U}\,  \phi_{\sigma\tau 1} \\				
				+\, \frac{A_\perp }{C_\perp - M_\perp s_\tau} \,
				\phi_{\bar\sigma\bar\tau 2}
			&=&  k  \,
			\phi_{\sigma\tau 2}\; ,
		\end{array}
		\right.
		\label{eq11}
\end{equation}
where we have omitted the spatial dependence of the wave function components.

In (\ref{eq11}) is a non-Hermitian eigenvalue problem yielding the wave numbers $k$ as eigenvalues. Wave numbers can be complex, with vanishing or 
nonvanishing imaginary parts, as required to include the possibility of both propagating and evanescent modes. The approach can be understood as an inverse problem where we fix $E$ and determine the complex wave numbers $k$.

\begin{figure}[tb]
	\centering
	\includegraphics[width=0.7\linewidth]{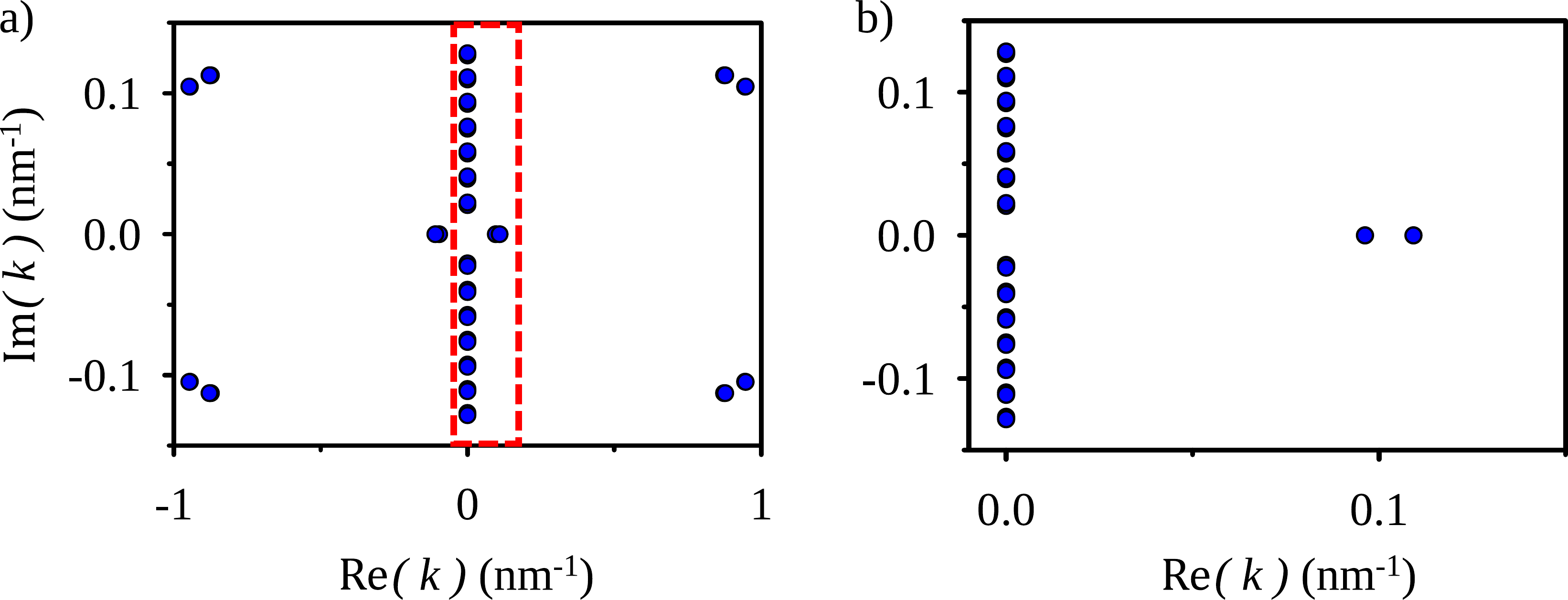}
	\caption{
		a) Wave numbers of the eigensolutions in the central section of the 3D~TI nanowire considered in figure~\ref{F4}.
		b) Close-up of a) on the two real wave numbers of which the degeneracy is lifted by the magnetic field, which is responsible for the beating pattern of the conductance in figure~\ref{F4}.
	}
	\label{FA1}
\end{figure}

In practice, we diagonalize the matrix resulting from (\ref{eq11}) using the ARPACK library which allows an efficient treatment for large sparse matrices~\cite{arpack}. An illustrative example of complex wave numbers obtained from (\ref{eq11}) is shown in figure~\ref{FA1}. The modes are distributed in a symmetric way in the complex plane, taking positive and negative values for real and imaginary parts and, for the particular case of figure~\ref{FA1}, there are 4 modes on the real axis representing the propagating modes.

\end{document}